# University-Industry-Government Relations in China:
# An emergent national system of innovations

Loet Leydesdorff [•] & Zeng Guoping [*]

**Abstract**

Since 1992, a new Chinese innovation system has been emerging in terms of university-industry-government relations. In recent years, science parks, incubators, and high-tech development zones have been provided with strong incentives. The commitment of the Chinese government to the further introduction of a market economy has been elaborated with a focus on the knowledge-base. The model of institutional adjustments has been replaced with systemic and evolutionary thinking about niche development and human resource management. Government interventions remain important for interfacing knowledge-based developments with those on the market. For example, new legislation on issues of "intellectual property rights" is crucial to the absorption of venture capital into these developments.

[•] Science & Technology Dynamics, University of Amsterdam, Faculty of Social and Behavioural Sciences, OZ Achterburgwal 237, 1012 DL Amsterdam, The Netherlands; Tel.: +31-20-525 6598; Fax: +31-20-525 2086; loet@leydesdorff.net; http://www.leydesdorff.net/

[*] Center of Science, Technology and Society, Tsinghua University, Beijing, 100084, China; Tel.: +86-10- 6278 1875; Fax: +86-10- 6278 4663; zenggp@mx.cei.gov.cn;
http://www.tsinghua.edu.cn/docsn/rwxy/stsweb/index.html

**Introduction**

In a comprehensive collection of comparative studies of "national innovation systems," Richard Nelson (1993)[1] did not include a chapter about Mainland China. Although the construction of a national innovation system in the People's Republic can be traced back to 1978 (when the Cultural Revolution ended), major decisions about the establishment of a market economy 'system' were not taken until the 14th National Conference of the Chinese Communist Party in 1992.

While the period of 1985-1992 can be characterized in terms of "structural adjustments", the deliberate decision "to rejuvenate the country through science and education" implied the use of evolutionary systems thinking about human resource management and the restructuring of institutional relations among universities, industry, and government.[2] The expansion and commercialization of the Internet after 1995 boosted the modernization process. Universities play an increasingly important role in the transformation of Chinese society towards a knowledge-based economy.

**Start-up firms**

"We started our enterprise in August 1998 when we were ten graduate students at Tsinghua University. However, we all graduated in 1999; two of us then left for the U.S. and two other students joined the company. Nowadays (that is, August 2000), we have 61 employees and we expect to have net sales of more then 10 million Yuan[1] during the fiscal year 2000. Our capital assets have grown from 0.5 million Yuan to over two million."

Huan Yu Xun is one of the ten students who created Beijing Smartdot Tech. Co., Ltd. Nowadays, he sees the fast rate of growth of his company as its major problem. As an SME located within the Science Park of the prestigious Tsinghua University, the company is still able to hire Master's and Ph.D. students. But the success is so great that the firm will soon outgrow the buildings of the Science Park. (They are currently paying a rent of Yuan 8400 / month to the Science Park.) What will happen if this growth continues in the next few years is almost inconceivable. They already have to rethink the whole enterprise, and they may have to become much more business-oriented (as opposed to knowledge-oriented).

China suffered from the Asia crisis precisely during the period of the ascent of such SMEs as Smartdot. Two factors, however, have mitigated the crisis. Unlike the other Asian economies, the Chinese Yuan is not yet completely convertible so that the government has kept a handle on speculative capital transfers. Second, the provision of venture capital—although available—is often so difficult that these ten students decided to invest Yuan 50,000 each in their own company by borrowing money from family and friends.

These advantages and disadvantages are typical for the Chinese situation. Do these conditions provide a competitive edge? On the one hand, the state is present much more prominently than in the West, since the central government itself orchestrates "the transition to the market economy." During the last few years, the means of this orchestration have included the development and exploitation of the knowledge capacities in the human resource base. Both public research institutes and universities have increasingly been functionalized to stimulate the economic development of the country.

---

[1] US$ 1 is approximately equal to Yuan 8

On the other hand, the political system itself had to be further developed to accommodate the requirements of the complex dynamics of a knowledge-based economy. As Wang Hongjia, a director of the new Zongguancun Science Park in Beijing, explains: the Chinese government has supplied two billion Yuan of venture capital to the Park, but most of the money has not yet been used. Although this money can be used, it cannot be owned, and therefore the efficiency of the transfer is often complicated.

Intellectual property rights are one of the major issues in knowledge-based innovation systems everywhere. But in China—as against the older market economies—there are no established practices for handling the intricacies involved in the transfer of knowledge.[3] "Who owns what?" is difficult to determine in the case of the failure of a new business and also in the event of its success. Which part should go to the university? While universities in the U.S. and Western Europe have experience in such cases,[4] legislation is lagging behind in the PRC? The exemption from tax for start-up firms (which remains the main instrument for stimulating SMEs) is valid only for three years. Although start-ups can profit for another three years from a reduced rate, in some sectors it is difficult to make a profit even within a period of five or six years.

The success rate of the SMEs in the Beijing area is estimated by our informant (Wang Hongjia) at 30 percent, that is, 600 high-tech corporations during the last ten years, in the northwest corner of the Beijing area where the main center of Chinese Academy of Science and more than ten universities are located (including the two major universities, that is Peking University and Tsinghua University). The mere size of the operation of this so-called "high-tech development zone" is beyond imagination of the European visitor.

The total area of the Zongguancun Science Park contains 39 member universities. The number of employees is now over 200,0000 and still increasing rapidly. According tot the statistics of 1998, there were 173,000 employees, of whom 7.6% had obtained Ph.D. and Master's degrees, 35.7% Bachelor's degrees, and another 22.0% had graduated from junior colleges and technical secondary schools. The whole enterprise is deeply rooted into the scientific community by deliberate policies. Of the 383 members of the Scientific Board, 300 are based in the Academy of Science and the Academy of Engineers, that is, they are among the top-level scientists and engineers of the People's Republic.

The development of the Zongguancun Science Park which integrates five existing science parks in the Beijing region is a recent development. The formal decisions were taken only in 1999, and most of the budget is not yet used in the operations. The management centre is not even located on the spot, since the office buildings are still under construction. But the development is managed on-the-fly, as is now more common in the Chinese innovation system. The "high-tech development zones" are constructed on the basis of principles similar to the "free zones" which were functional in opening the Chinese economy during the first half of the 1990s. The latter were business-oriented, while the new developments are knowledge-intensive.

Fifty-three "high-tech development zones" have now been established throughout the country. Like the Zongguancun Science Park, they can count on government support in order to solve problems at the interfaces between the economic forces of the market, legislation, and knowledge input. The Chinese government has chosen the role of making the political system supportive of the introduction of a market economy and a knowledge-based society. This role is consistent with the professed role of the communist party as the main agent of change and modernization in an otherwise still developing country. Yet, the obstacles may be as big as the opportunities.

**Historical developments**

One can distinguish different phases in the recent history of the political economy of China. After the Cultural Revolution, the first phase (1978-1985) emphasized the *institutional* reconstruction of the science & technology system. The focus in public policy shifted only gradually away from regarding the "class struggle as the central task" to a programme of "modernization" and "economic construction". The reconstruction of an S&T system and a focus on human resource management (e.g., higher education) were accordingly central to this period of institution building. Many research institutions which are still active, were founded in the 1978-1985 period.

The second period (1985-1991) can be characterized in terms of structural and institutional *adjustments* within the S&T system to the exigencies of market forces.[5] Among other things, competition was introduced and institutions were increasingly evaluated and changed to improve their performance. The High-tech Research and Development Program (Program 863) was launched in 1986. The first incubator thereafter emerged in 1987 in Wuha. The national high-tech development zone—the Beijing New-tech Development Trial Zone (which functioned as the predecessor of the Zhongguancun Science Park)—was launched by the government during this period of institutional reforms.

New relations (e.g., liaison offices) were established in the inter-institutional sphere, but the system of institutions and institutional relations had not yet been challenged in terms of its capacity to respond to new political and economic goals. After the Tienanmen Square incidents and, more generally, after the hampering of the reforms in the period 1988-1991, a more revolutionary transformation was proclaimed in 1992. This reform was formulated in terms of "*state-coordinated, industry-dominant*." The functions of the state and its apparatuses should themselves now be reshaped to support the market-economic development of the country. For example, 242 research institutes which belonged to the various ministries, have now changed their status into enterprises. Only last year, about twenty new science parks were certified at university campuses by the government. The goal is to set up more than one hundred university science parks all over the country in the near future.

One should note the differences from the Western model. The Chinese approach is always systemic, and the mere size of the operation is astounding because of the rate of replication available to relative successes. The function of the national government remains central, even in the process of devolution. The transition is considered to require political intervention and new legislation continuously, since the old institutions cannot be expected to change their functions without resistance. Thus, the state remains an important agent of change. However, the mode of operation has changed from a top-down and *ex ante* planning to a bottom-up receiving of signals from the market forces and an *ex post* regulation whenever government intervention seems necessary.

The strong state has the potential advantage of regulation. Among other things, one wishes to prevent the emergence of extreme poverty, as, for example, in Latin America. The Chinese population seems to enjoy the consumerist values which are now replacing the ideals of an egalitarian paradise. Already during the period of the experiments with "free zones", neo-evolutionary concepts like "niche management" and "human resource management" were actively used in the politically-guided development of new business activities, for example, on the eastern shore around Hong Kong and Shanghai.[6] Evolutionary theorizing (Marx, Schumpeter) facilitated the idea of considering the dynamics as complex in terms of modes of operation that can be differentiated functionally. From this perspective, the main problems have to be solved at the relevant interfaces.

The adjustment process and the restructurings require contributions from the academic community, but the changes in the latter's functions feed back onto the carrying agencies. Roles have continuously to be redefined with respect to further development. One major problem, for example, has been the devolution of the (large) Public Research Institutes. After a period of stabilization of budgets, these institutes were increasingly given the task of "facing the market", for example, by making a substantial percentage of their (lump-sum) budget conditional on their commercial success. Thus, the transformation process no longer focuses on changing the functional roles of existing institutions, but on shaping an innovation system with functional components into a "state-coordinated, industry-dominant" mode of production.

**Towards a knowledge-based economy**

Perhaps China has been extremely fortunate in taking these steps in the period when the Internet was emerging. Although the Internet also created political problems—for example, by allowing users free access to Western sources of information, while CNN and BBC are still not included in the cable packages of large cities like Beijing and Shanghai—the combination of Intranet en Internet resources has provided a wealth of opportunity for Chinese research, development, and business. The above-mentioned firm of Smartdot, for example, has specialized in knowledge management among databases at the interfaces between Intranet and Internet configurations.

During this process—driven by the government—the transformed institutions gain momentum and are then provided with increasing autonomy. The idea is to make the sources of finance multi-channel and multi-level so as to increase the internal complexity of the organization, i.e., its buffering capacity in rapidly changing environments, as well as its knowledge-intensity and potential contribution to further modernization. Although government expenditure in R&D decreased as a percentage of GNP during most of the 1990s, the total volume of R&D has nevertheless increased considerably.

The knowledge-base of the economy is prominently present. It is visible in the numerous new buildings which are constructed in the neighbourhood of university campuses. But it is also present in policy formulations. The intellectual resource base is to be combined with the ingenuity of Chinese businessmen on the markets in order to give the economy a leading edge. While the ten entrepreneurs of Smartdot at Beijing identify with professional values more than with commercial ones, the explosion of commerce and consumerism in an otherwise developing society is most impressive in the new business areas in Pudong, across the Hangpo river in Shanghai. University-industry-government relations are challenged to keep this development progressive and balanced.

**Conclusions**

The construction of a Chinese innovation system has been favoured by the increasing knowledge-intensity of the economy, the internal resources of the academic system (e.g., the sheer number of academicians, students, and intellectuals), the return from abroad of (especially U.S.) academics, and the firm commitment of the national government to the creation of a market economy. The deliberate coupling of the legislation system to the functional requirements of a knowledge-based economy has made it possible for the authorities to postpone more difficult political debates concerning further democratization, the unification of China, etc. by focusing on economic reforms. One prefers to begin with the problems which are most likely to prove soluble.

This policy has been successful in the case of the unification with Hong Kong, but it remains to be seen whether a system with such an established strong overlay of university-industry-government relations will ever be able to unite with the more entrepreneurial mode of production of Taiwan. From a Western perspective, one would like to see the further development of a pluriform and multi-party democracy. Yet, what we are witnessing nowadays in China, in our opinion, is the most important transition of a political economy ever performed under the conditions of peace. The transition is not without coercion, but there is no threat of civil war or instability in the society which might lead to a *coup d'état*, etc. The transformation is firmly guided by the communist state, but in a social-democratic mode. It seems to us that this can be considered as a major achievement.